# LOT-G3: LÂMPADA DE PLASMA, OZONIZADOR E TRANSMISSOR CW
*LOT-G3: Plasma Lamp, Ozonator and CW transmitter*

**Ricardo Gobato** [ricardogobato@seed.pr.gov.br]
*Secretaria de Estado da Educação do Paraná (SEED/PR)*
*Av. Maringá, 290, Jardim Dom Bosco, Londrina/PR, 86060-000, Brasil.*
**Desire Francine Gobato** [desirefg@bol.com.br]
Cinemark
*Av. Theodoro Victorelli, 150, Boulevard Londrina Shopping, Jardim Interlagos, Londrina/PR, 86027-750, Brasil.*
**Alekssander Gobato** [alekssandergobato@hotmail.com]
*Faculdade Pitágoras Londrina*
*Rua Edwy Taques de Araújo, 1100, Gleba Palhano, Londrina/PR, 86047-500, Brasil.*

**RESUMO**

O LOT-G3 foi desenvolvido para se um equipamento versátil, que realize vários experimentos simples para serem utilizados no auxílio às aulas de física para o ensino médio. De fácil construção, baixo custo, utiliza materiais de fácil acesso. Sua construção envolve práticas simples e conhecimento de eletromagnetismo. Tem a função de um globo de plasma, para demonstrar à ionização de um gás a baixa pressão, bem como a formação de campo magnético. Pode ser utilizado como higienizador de ambientes fechados, como veículos automotores na função ozonizador, demonstrando a ionização do oxigênio presente na atmosfera, produzindo ozônio, essencial a vida na terra. E como um transmissor faiscador, de baixa potência, baixa frequência em amplitude modulada em onda contínua (CW), para sinais em código Morse. Portanto o equipamento aqui denominado de LOT-G3, apresenta três funções: uma lâmpada de plasma, ozonizador e um transmissor CW.

**Palavras-chave**: código Morse; ozônio; plasma; transmissor CW.

**ABSTRACT**

The LOT-G3 is designed to be a versatile equipment that perform several simple experiments for use in helping the physics classes for high school. Easy construction, low cost, using easily accessible materials. Its construction involves simple practices and knowledge of electromagnetism. It has the function of a plasma globe to demonstrate the ionization of a low pressure gas, as well as the formation of magnetic field. Can be used as sanitizer closed environments such as automotive vehicles in ozonator function, demonstrating the ionization of oxygen in the atmosphere, producing ozone, essential to life on earth. And as a sparks transmitter, low power, low frequency modulated continuous wave in (CW), for signals in Morse code. Therefore the equipment here called LOT-G3, has three functions: a plasma lamp, ozonator and CW transmitter.

**Introdução**

O objetivo inicial do experimento foi construir um globo de plasma (GP) utilizando materiais de fácil acesso e baixo custo, para ser utilizado no ensino de física para o ensino médio, demonstrando a ionização de um gás a baixa pressão, bem como o campo magnético criado. Para aumentar a intensidade do campo magnético induzido no entorno do GP, utilizou-se duas voltas e meia de fio sólido de alumínio, envolta da lâmpada, formado um solenoide. Como GP utilizou-se uma lâmpada incandescente comum, com filamento intacto ou rompido, Como esse se mantém em contato parcial com o vidro da lâmpada, formou-se externamente plasma, que produz forte ionização do gás no seu entorno devido à diferença de potencial do globo e do solenoide, produzindo grande quantidade do ozônio, do qual é percebido rapidamente pelo forte odor logo no início de seu funcionamento. Teve-se então a ideia de ampliar a função do equipamento acrescentando-se um solenoide, ao lado da lâmpada com uma pequena distância, do qual esta sofreria a indução do campo magnético da lâmpada-solenoide (LS), formando uma corrente de indução do campo eletromagnético variável da LS. Sendo a bobina ligada a uma antena, ou mesmo sem esta, deixando somente um pequeno fio solto. Teremos então agora uma lâmpada de plasma, um ozonizador e um transmissor de CW, em um único equipamento, denominado de LOT-G3.

**Fundamentos**

Plasma

Pode-se dizer que com o desenvolvimento da energia elétrica e a descoberta da estrutura do átomo levaram à descoberta de um novo estado da matéria. Mais de 100 anos desde Sir William Crookes em seus experimentos com descargas elétricas em tubos de vácuo, em 1879, o chamava "matéria radiante". Sugeriu então a ideia da existência de um novo tipo de gás: um gás composto de partículas carregadas, tais como uma mistura de um gás de elétrons e um gás de prótons. Em meados do século 19 o fisiologista checo Jan Evangelista Purkinje uso o termo plasma, do grego que significa "formado ou moldado" para designar o líquido claro que permanece após a remoção de todo o material corpuscular no sangue. A natureza da matéria do "raio catódico" do tubo de Crookes foi depois identificada pelo físico britânico Sir J.J. Thomson em 1897, e mais tarde, o químico americano Irving Langmuir investigou as descargas elétricas em gases e propôs em 1922 que os elétrons, íons e neutros em um gás ionizado possam ser igualmente considerados como material corpuscular arrastado em algum tipo de meio fluido e em 1929 juntamente com outro cientista americano, Levy Tonks, usou o termo plasma para descrever as oscilações da nuvem de elétrons durante a descarga. Esta nuvem de elétrons brilhava e "balançava", à semelhança de uma substância gelatinosa que lembrou Langmuir de um plasma sanguíneo. No entanto, o termo plasma nas experiências de Langmuir eram completamente equivocadas. Plasma em física designa-se como um fluido condutor constituído por uma mistura de um gás de elétrons, íons e átomos neutros, enquanto que o plasma no sangue é o líquido de cor amarelada clara, em que as células do sangue são carregadas por este líquido. [Eliezer e Yaffa, 2001; Bellan, 2004; Wikipédia, 2015]

O plasma é considerado o quarto estado físico da matéria, similar ao gás, no qual certa porção das partículas é ionizada. O fundamento básico para a formação de plasma é que o aquecimento de um gás provoca a dissociação de suas ligações moleculares, levando seus átomos constituintes à ionização deste gás. A presença de um número considerável de portadores de carga torna o plasma eletricamente condutor, de modo que este responde fortemente a campos eletromagnéticos. Como o gás, o plasma não possui forma ou volume definido, a não ser quando contido em um recipiente; sob a influência de um campo magnético ele pode formar estruturas como filamentos, raios e camadas duplas, etc. Alguns plasmas comuns são as estrelas e placas de neônio. No universo, o plasma é o estado mais comum da matéria, cuja maior parte da qual se encontra no rarefeito meio intergaláctico e nas estrelas. [Eliezer e Yaffa, 2001; Bellan, 2004; Wikipédia, 2015; Braithwaite e Chabert, 2011]

Globo de Plasma

O globo de plasma ou lâmpada de plasma é essencialmente constituído por uma esfera de vidro ou acrílico com um gás a baixa pressão e por um eletrodo central a alta voltagem. Descargas elétricas provocam a excitação e a ionização de alguns átomos de gás. Os átomos excitados, ao voltarem ao estado inicial, emitem luz. Em uma lâmpada fluorescente os filamentos aquecidos são capazes de "excitar" os elétrons do gás a baixa pressão, ionizando o gás, geram uma descarga elétrica. Em uma lâmpada de plasma a alta tensão ionizando o gás, rompe o meio dielétrico, fazendo-o passar para o estado plasma, o tornado condutor. Quando uma pessoa coloca a mão na lâmpada acima da zona iluminada, ela ilumina até à zona em que a mão encosta, pois a pessoa passa a ser o condutor elétrico, induzindo a corrente à área onde a mão está. [Eliezer e Yaffa, 2001; Bellan, 2004; Wikipédia, Plasma, 2015].

Das formas comuns de plasma têm-se os produzidos artificialmente, como as telas de plasma, lâmpadas fluorescentes, coroas geradoras de ozônio, etc.; os plasmas terrestres, como os raios, fogo de Santelmo, auroras polares, o fogo fátuo, etc.; e os espaciais, como as estrelas, o vento solar, meio interplanetário, interestelar e intergaláctico, nebulosas, etc. [Wikipédia, Plasma, 2015; Braithwaite e Chabert, 2011].

Radiofrequencia

Em termos gerais, a tecnologia de radiofreqüência (RF), ou transmissão-recepção sem fio, é a utilização de ondas electromagnéticas do espectro entre 3 Hz e 300 GHz. É sem dúvida uma das tecnologias mais importantes na sociedade moderna. A possibilidade de ondas eletromagnéticas foi postulada pela primeira vez por James Clerk Maxwell em 1864 e sua existência foi verificada por Heinrich Hertz em 1887. Nikola Tesla foi o primeiro a demonstrar a transmissão sem fio por rádio, já com tecnologia de controle remoto [Trinkaus, 1988; Tesla, 1909; Tesla, reimpressão Fredonia Books, 2002]. Em 1895, Guglielmo Marconi tinha demonstrado o rádio como uma tecnologia de comunicação eficaz. Com o desenvolvimento da válvula termiônica, no final do século XIX, a tecnologia de rádio se tornou um meio de comunicação e entretenimento de massa. A primeira metade do século XX viu desenvolvimentos, tais como radar e televisão, que alargava o âmbito desta tecnologia. Na segunda metade do século XX, grandes avanços vieram com o desenvolvimento de dispositivos semicondutores e circuitos integrados. Esses avanços tornaram possível a redução das dimensões dos dispositivos de comunicação, tornando-os compactos e portáteis, o que resultou na revolução das comunicações móveis. O tamanho dos componentes eletrônicos continua a diminuir e, em consequência, novas áreas inteiras com a nanotecnologia sugiram. Em particular, as comunicações de espalhamento espectral em freqüências gigahertz são cada vez mais utilizadas para substituir cabos e outros sistemas que fornecem conectividade local. [Coleman, 2004]

James Clerk Maxwell

O trabalho de Maxwell, e outros pioneiros, levaram ao desenvolvimento de equações que descrevem totalmente fenômenos electromagnéticos. Essas equações revelam uma ligação íntima entre os campos elétricos e magnéticos e mostram que há circunstâncias em que um campo não pode existir sem o outro. Em particular, as equações que preveem a existência de ondas electromagnéticas e demonstram que, para um dado campo de onda elétrico, existe um campo magnético associado. Hertz em 1887 comprovou a existência de ondas eletromagnéticas e também a propriedade de que a amplitude da onda decai como o inverso da distância da fonte. Esta lenta decadência em amplitude é a propriedade que faz com que as ondas eletromagnéticas um meio eficaz de comunicação, como demonstrado por Marconi, no final do século XIX. Infelizmente, as ondas electromagnéticas têm uma estrutura muito mais complexa do que muitas outras ondas (ondas sonoras, por exemplo) e esta complexidade deve ser compreendida para ser utilizada com sucesso [Coleman, 2004]. Como consequência, uma compreensão de ondas eletromagnéticas deve ser entendida e a seguir são apresentadas as equações de Maxwell no seu formalismo diferencial. [Maxwell, 1878]

$$\vec{\nabla} \cdot \vec{E} = \frac{1}{\varepsilon_0} \rho \qquad (1),$$

$$\vec{\nabla} \cdot \vec{B} = 0 \qquad (2),$$

$$\vec{\nabla} \times \vec{E} = -\frac{\partial}{\partial t} \vec{B} \qquad (3) \text{ e}$$

$$\vec{\nabla} \times \vec{B} = \mu_0 \vec{j} + \mu_0 \varepsilon_0 \frac{\partial}{\partial t} \vec{E} \qquad (4)$$

Têm-se as constantes, $\mu_0 \varepsilon_0$, e $c$. Onde $\mu_0 \varepsilon_0 c^2 = 1$, e $c = 2{,}299792458.10^8 \, m.s^{-1}$ é a velocidade da luz no vácuo. $\mu_0 = 4\pi.10^{-7} \, N.A^{-2}$ a permeabilidade magnética no vácuo e a permissividade elétrica no vácuo $\varepsilon_0 = 8{,}854187817.10^{-12} \, C^2 N^{-1} m^{-2}$.

Nas equações acima $\vec{\nabla} = \hat{i}\frac{\partial}{\partial x} + \hat{j}\frac{\partial}{\partial y} + \hat{k}\frac{\partial}{\partial z}$ é o operador diferencial *nabla*, $\rho$ é a densidade de carga e $\vec{j}$ é a densidade de corrente elétrica. [Maxwell, 1878]

A equação (1) é conhecida com a lei de Gauss para a eletrostática, que estabelece uma relação entre o fluxo de campo elétrico através de uma superfície fechada e as cargas que estão no interior dessa superfície. A equação (2) é conhecida com a lei de Gauss para o magnetismo, que estabelece o fluxo de campo magnético através de uma superfície fechada como sendo o divergente da componente do campo magnético normal à superfície é igual à zero, implicando a não existência de monopolos magnéticos. A equação (3) é conhecida com a lei de Faraday que estabelece que a força eletromotriz induzida em um circuito fechado é determinada pela taxa de variação do fluxo magnético que atravessa o circuito e a equação (4) é a lei de Ampère-Maxwell, que descreve a produção de campos magnéticos por corrente elétrica.

Onda Contínua

Rádio AM é o processo de transmissão através do rádio usando Amplitude Modulada (AM). É transmitido em várias bandas de frequência. Foi por oitenta anos o principal método de transmissão via rádio. É caracterizada pelo longo alcance dos sinais, mas frequência AM está sujeita a interferências de outras fontes eletromagnéticas. [Wikipédia, Rádio, 2015]

O CW (*continuos wave*) onda contínua também é o nome que se dá a um antigo método de radiotransmissão, no qual uma onda portadora é "ligada" e "desligada". Nas antigas radiotransmissões da telegrafia wireless, as CW eram utilizadas para transmissão-recepção com o uso é o Código Morse. As transmissões de rádio usaram transmissores de faíscas para produzir oscilações de rádio frequência na antena de transmissão; esses sinais apresentavam uma característica de atenuação de amplitude durante cada pulso de energia radiada caracterizando essa técnica como emissão de ondas "contínuas". Enquanto a transmissão e recepção não permitia a transmissão de áudio da forma que é feito pela Modulação em amplitude com a complexidade do áudio atual, o CW era a única forma de comunicação por radio disponível no inicio. [Wikipédia, Rádio, 2015]

O CW continua sendo usado mesmo tendo a comunicação com voz se tornando perfeita. Devido a baixa área de frequência do sinal que possibilita o CW transpor longas distancias com condições de propagação de onda, onde a modulações em AM e voz se perderiam. Um simples transmissor de meio watt desse tipo de transmissão de baixa potência o CW pode transmitir milhares de quilômetros com condições de propagação. No radioamadorismo, os termos "CW" e "código Morse" são frequentemente usados como se fossem sinônimos, apesar das distinções entre os dois (o código Morse pode ser emitido via sons ou luzes, por exemplo). [Wikipédia, Radio, 2015]

Os transmissores de CW são simples e baratos, e o sinal CW transmitido não ocupa muito espaço em frequência (geralmente, menos que 500 Hz). Entretanto, os sinais de CW serão difíceis de serem ouvidos em um receptor normal; você irá ouvir apenas um rápido e fraco período onde o ruído de fundo se torna quito conforme os sinais CW são transmitidos. Para superar este problema, os receptores de radio amadores e de ondas curtas incluem um circuito oscilador de frequência de batimento denominados *BFO*. A transmissão feita por CW é a mais simples forma de modulação. A saída do transmissor é chaveada, ou seja, ligada e desligada, tipicamente para formar os caracteres do código Morse. [Tipler e Mosca, 2012]

Ozônio

Em 1785 o químico holandês Martinus Van Marum conduzia experimentos envolvendo faíscas elétricas acima da água quando notou um cheiro estranho, que ele atribuiu às reações elétricas, deixando de perceber que ele tinha de fato criado o ozônio ($O_3$). Meio século mais tarde, Christian Friedrich Schönbein notou o mesmo odor pungente e reconheceu-o como o cheiro que muitas vezes ocorria na sequência de um raio. Em 1839, ele conseguiu isolar o produto químico gasoso e nomeou-o de ozônio, da palavra grega *ozein* (ὄζειν) que significa cheirar. [Rubin, 2008; Soret, 1865; Kogoma e Okazak, 1994]

O $O_3$ ou trioxigênio é um alótropo triatômico do oxigênio muito menos estável que o diatômico $O_2$. Forma-se quando as moléculas de oxigênio $O_2$ se rompem principalmente devido à radiação ultravioleta que vem do sol, e os átomos separados combinam-se individualmente com outras moléculas de $O_2$. Em CNTP é um gás de tonalidade azul, massa molecular 47,998g/mol, ponto de fusão 80,7K, densidade 2,144G/L, ponto de ebulição 161,3K, solubilidade em água 0,105g/100 ml (273K) e índice de refração 1,226 no estado líquido. [Streng, 1961]

O $O_3$ um poderoso agente oxidante, em segundo lugar entre os elementos atrás apenas do flúor. Pode oxidar muitos compostos orgânicos e é utilizado comercialmente como um branqueador de ceras, óleos e produtos têxteis, e como um agente de desodorização. Sendo um poderoso germicida, é também utilizado para esterilizar o ar e a água de beber. O $O_3$ é geralmente fabricado pela passagem de uma descarga elétrica através do gás $O_2$ ou através de ar seco, obtendo sempre uma mistura resultante de $O_3$, $O_2$ e ar. [Devins, 1956; Chen, 2002; Chemical of the Week, Ozone, 2015; UCDavis Chem, Ozone, 2015]

A ação benéfica do ozônio ao ser humano é bem conhecida. Mas seu efeito nocivo ou favorável à vida na Terra depende da altitude em que ele se situa. A maior parte desse gás está na estratosfera (entre 13 e 40 km), onde funciona como escudo à radiação ultravioleta. No entanto, próximo da superfície, o O3 é um poluente que causa danos ao tecido pulmonar dos animais e prejudica a vegetação. Na troposfera, sua ação chega a afetar o clima. Os relâmpagos, fenômeno natural que ocorre na Terra desde seus primórdios, contribuem para o aumento da concentração na atmosfera de compostos que podem ser nocivos ao homem, como o aumento da quantidade de $O_3$ e óxidos de nitrogênio (NOx) na atmosfera, próximos a superfície. [Devins, 1956; Chen, 2002; Kuck, 2003]

**Desenvolvimento**

A construção do equipamento LOT-G3 é simples, mas exige certa técnica. Como base foi utilizada uma caixa de polietileno de dimensões aproximadas de 17 x 11,5 x 11 cm, ou seja, um pote comum de sorvete de 2 (dois) litros, utilizados pelos fabricantes de sorvetes em geral. Uma lâmpada incandescente comum, de 60W, vidro transparente, podendo ter filamento rompido - "queimado" ou não ou seja, com ou sem defeito, e de qualquer marca, adquirida em supermercados, lojas de materiais de construção, etc. Um interruptor liga-desliga, 127V~220V, 5A, utilizado como interruptores de luz em casas de madeira, também podendo ser adquirida do mesmo modo anterior. Fios de cobre esmaltados obtidos, em lojas de materiais de construção. No LOT-G3 foi utilizado os fios obtidos de um pequeno transformador, do circuito eletrônico de uma lâmpada econômica fluorescente - lâmpadas econômicas utilizadas nos mais variados ambientes domésticos e comerciais. Um conector macho, 127V~220V, 5A. Três metros de fio duplo, paralelo, encapado, de espessura 2 mm, comum em redes elétricas residenciais, mas por economia foi aproveitada fiação de uma fonte AT de um microcomputador antigo, como também seus interruptores, *plugs* de conexão, parafusos, fios, etc. Bastão de silicone - "cola quente", para fixação, das L1 e L2, lâmpada, etc. Um tubo de papelão de papel higiênico, utilizado como base de enrolamento da bobina L1.

A tensão de alimentação do equipamento foi de 127V ou seja, a usina do fogão - transformador - utilizada foi de 127V, sendo a maioria bivolt, Figura 1. A tensão de saída para geração das faíscas é de aproximadamente 17KV, com baixa corrente 150mA. Dependendo do modelo, gera de 20 a 60 centelhas por segundo. Sendo uma usina de fogão 4 bocas, tem-se 5 faiscadores, contando com o forno. Para cada "queimador" do fogão existe uma saída exclusiva [Portal do Eletrodoméstico, 2012]. A lâmpada incandescente utilizada é confeccionada com filamentos, geralmente, feitos de tungstênio, A ampola é cheia com um gás inerte, para reduzir a evaporação do filamento e impedir a sua oxidação, a uma pressão de cerca de 70 kPa (0,7 atm). O gás é geralmente uma mistura de argônio (93%) e nitrogênio (7%), onde argônio é utilizado para a sua inércia, baixa condutividade térmica e baixo custo, e o azoto - nitrogênio é adicionado para aumentar a tensão de ruptura e evitar formação de arco entre as partes do filamento. [Wikipédia, Incandescent light bulb, 2012; Osram, 2015]

A lâmpada utilizada no experimento apresenta o mesmo efeito observado em uma bola de plasma, que é comumente demonstrado em feira de ciências, onde a bola de vidro contém ar a baixa pressão, a esfera metálica no centro está ligada a uma tensão elétrica alternada, de alta frequência e tensão, gerando o campo eletromagnético dentro da "bola" - globo de acrílico, onde os elétrons livres são acelerados e colidem com os átomos do gás e provocam a sua ionização, liberando mais elétrons, que também serão acelerados assim forma-se uma descarga elétrica no gás.

Na Figura 1 é vista a fotografia de uma usina para fogão de uso doméstico com vista para os terminais de tensão. A Figura 2 representa o esquema simples do circuito do equipamento denominado LOT-G3 (Lâmpada de Plasma, Ozonizador e Transmissor CW). Tem-se como legendas: M, manipulador telegráfico, ou simplesmente um interruptor comum de luz 127V~220V 5A; L1, bobina com núcleo de ar, com 48 mm de diâmetro, 100 mm de comprimento, fio de cobre esmaltado de 0,3 mm de espessura, com 300 espiras circulares; L2, bobina em alumínio, com 3 espiras circulares em fio sólido de alumínio de 2 mm de espessura, cujo núcleo é a lâmpada incandescente. Na Figura 3 é vista a fotografia do interior do LOT-G3, mostrando os terminais de ligação e o transformador (usina fogão quatro bocas), a base da lâmpada incandescente fixada em um pote de sorvete 2 litros. Na Figura 4 a fotografia do exterior do equipamento LOT-G3.

As Figuras 5 e 6 representam as fotografias de uma lâmpada incandescente comum [Osram, 2015], e de um globo de plasma, respectivamente. [Wikipédia, Globo de plasma, 2015]

A faixa de frequência abrangida pelo LOT-G3, compreende de 540 a 1600 kHz em AM, podendo ser captada por qualquer receptor comum, portátil, em veículos, residenciais, etc., com alcance aproximado de 100m com obstáculos, sem a utilização de antena externa. Um bom alcance tendo em vista a não utilização de antena e a baixa potência do LOT-G3.

**Figuras**

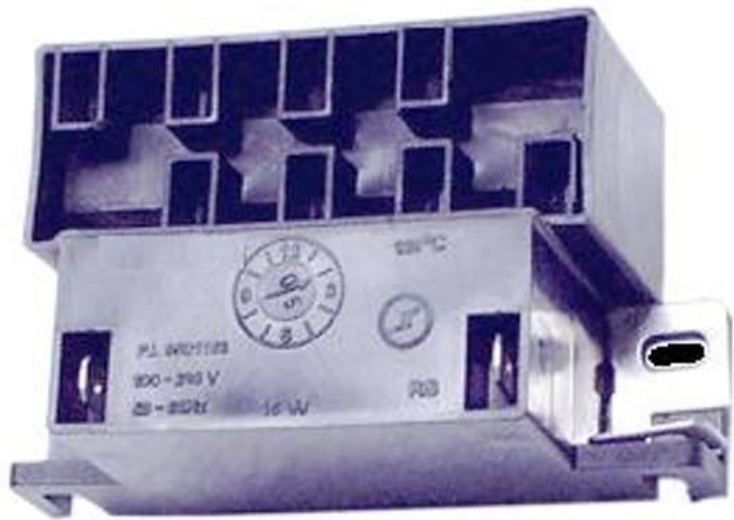

**Figura 1**. Fotografia de uma usina para fogão de uso doméstico com vista para os terminais de tensão.

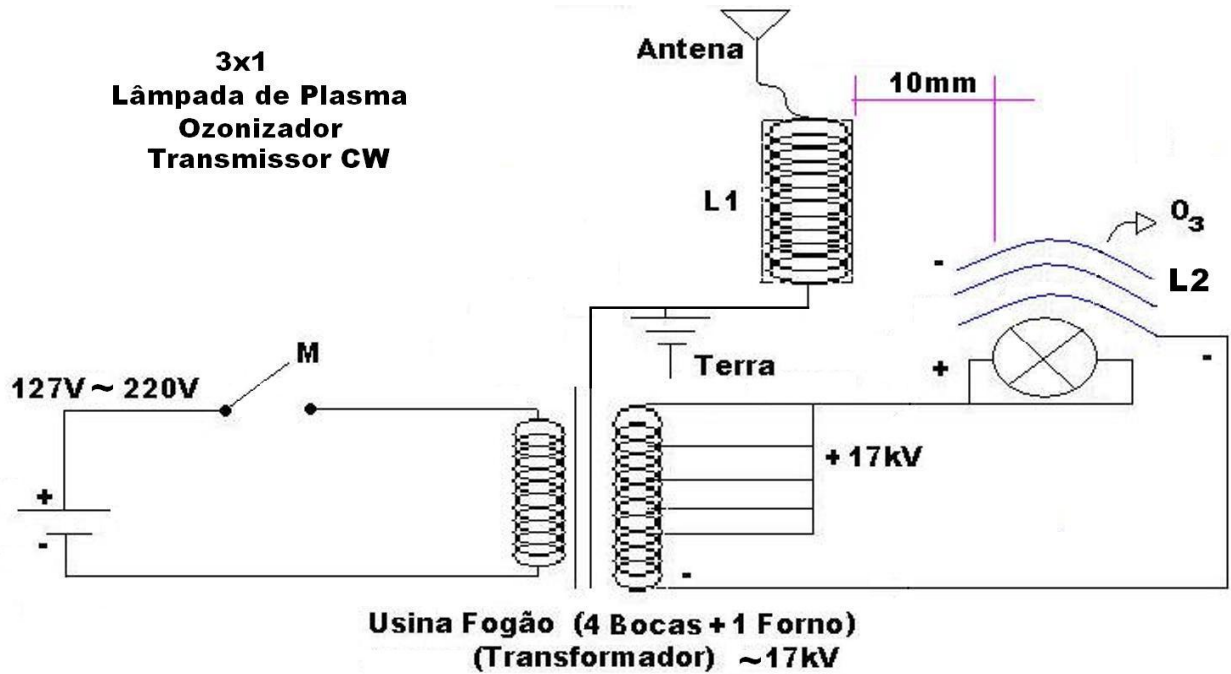

**Figura 2**. Esquema do circuito eletrônico do equipamento LOT-G3.

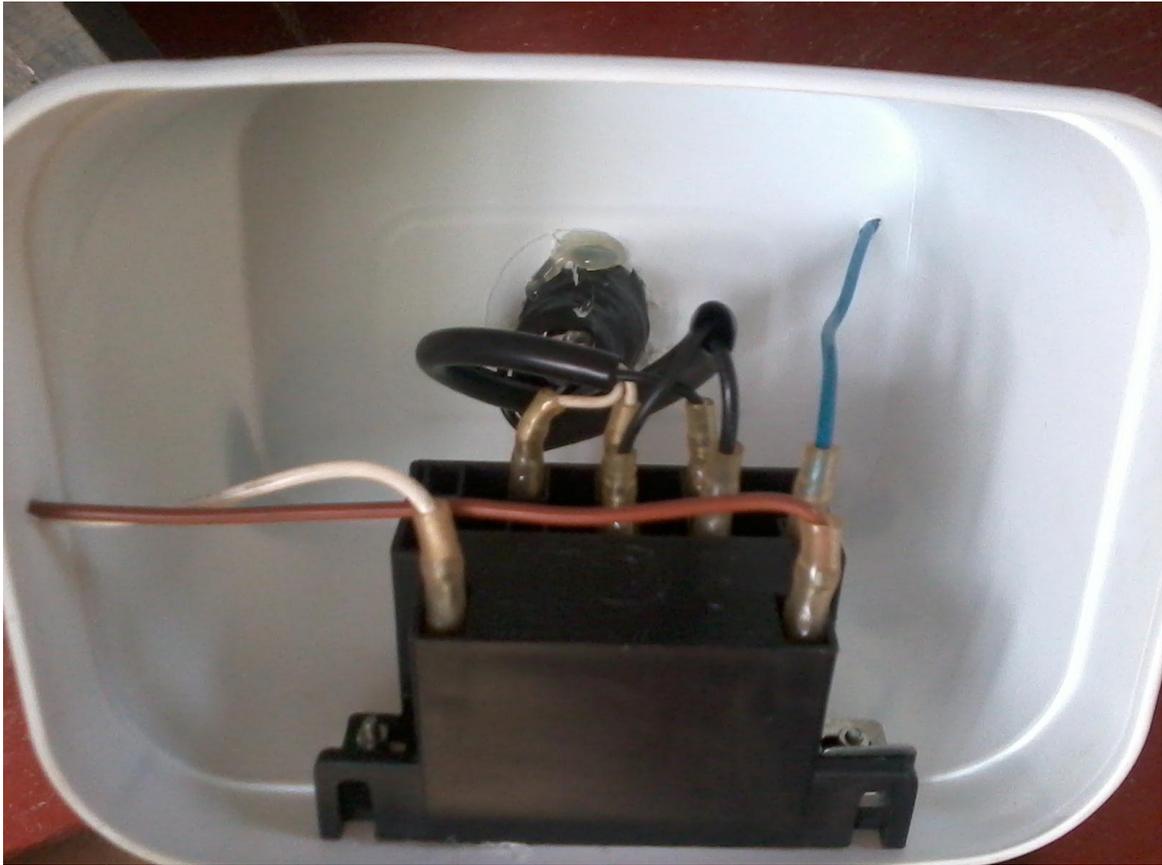

**Figura 3**. Fotografia do interior do equipamento LOT-G3, mostrando os terminais de ligação e o transformador (usina fogão quatro bocas), a base da lâmpada incandescente fixada em um pote de sorvete 2 litros.

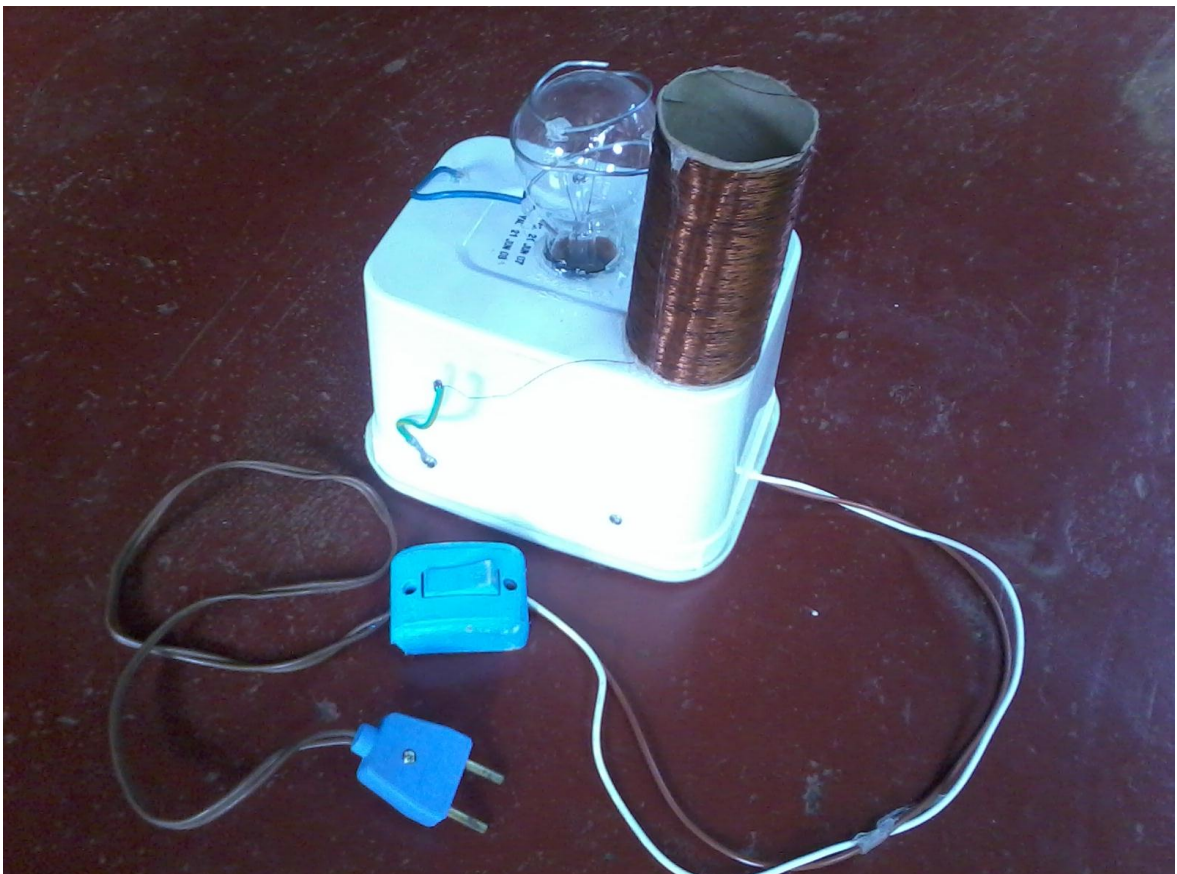

**Figura 4**. Fotografia do exterior do equipamento LOT-G3.

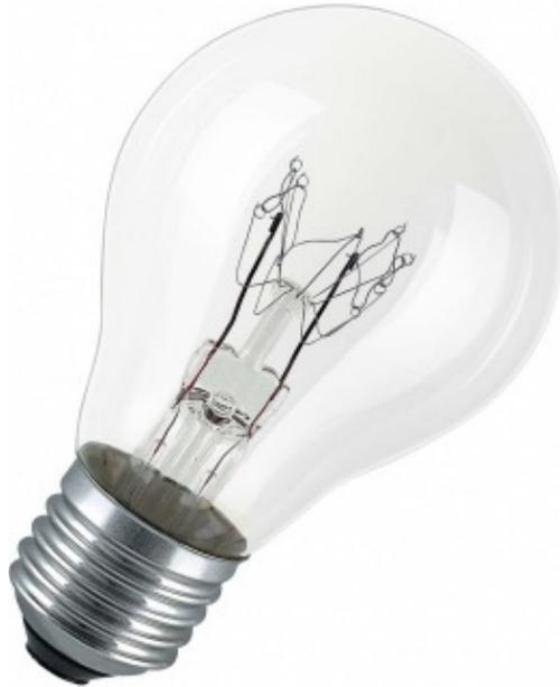

**Figura 5**. Fotografia de uma lâmpada incandescente comum. **Fonte**: [Osram, 2015].

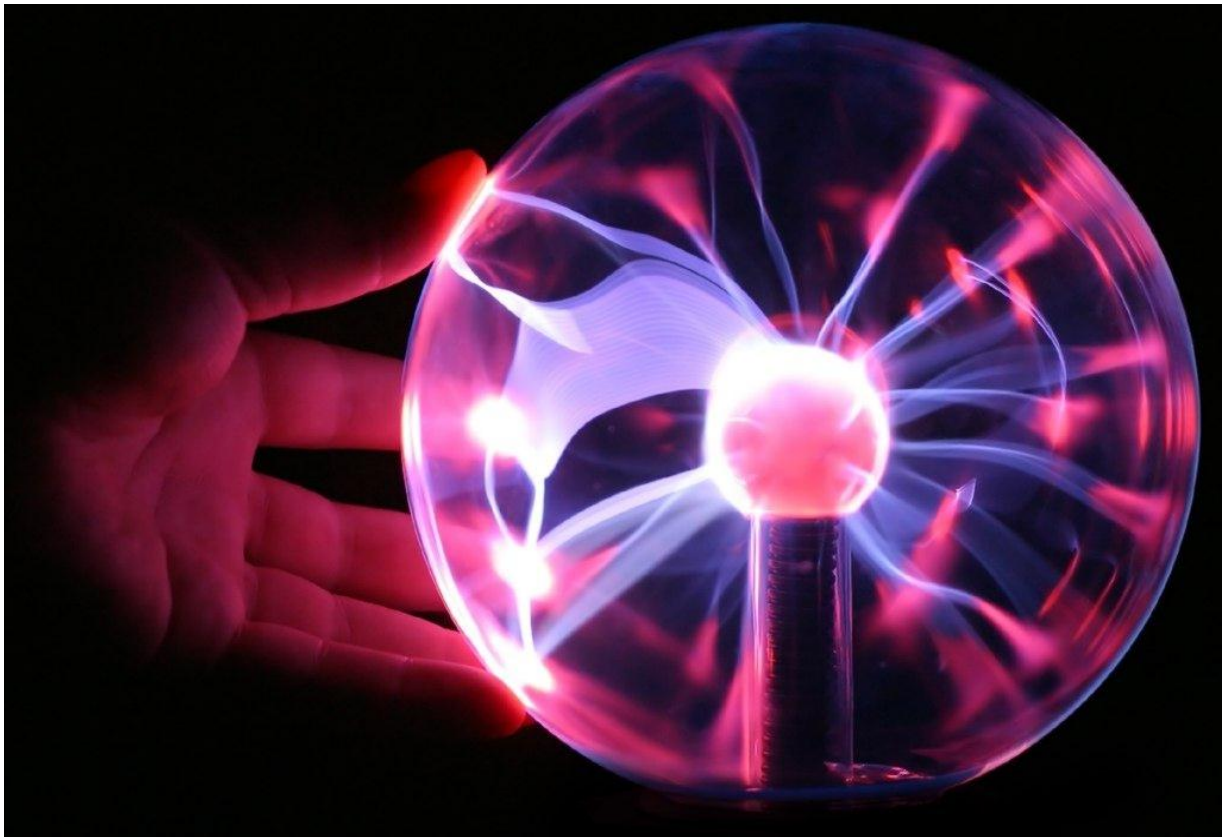

**Figura 6.** Globo de Plasma. **Fonte:** [Wikipédia, Globo de plasma, 2015].

### Conclusões

O LOT-G3 se presta muito bem para as funções do qual foi projetado e desenvolvido. Pode ser utilizado tanto em aulas de física para o ensino médio como ensino superior, em demonstrações de eletromagnetismo.